# An optimised multi-arm multi-stage clinical trial design for unknown variance

Michael J. Grayling[1], James M. S. Wason[1,2], Adrian P. Mander[1]

1. Hub for Trials Methodology Research, MRC Biostatistics Unit, Cambridge UK.
2. Institute of Health and Society, Newcastle University, Newcastle, UK

Corresponding author: Michael J. Grayling. Email: mjg211@cam.ac.uk. Address: MRC Biostatistics Unit, Cambridge Institute of Public Health, Forvie Site, Robinson Way, Cambridge Biomedical Campus, Cambridge CB2 0SR United Kingdom.

**Abstract**
Multi-arm multi-stage trial designs can bring notable gains in efficiency to the drug development process. However, for normally distributed endpoints, the determination of a design typically depends on the assumption that the patient variance in response is known. In practice, this will not usually be the case. To allow for unknown variance, previous research explored the performance of $t$-test statistics, coupled with a quantile substitution procedure for modifying the stopping boundaries, at controlling the familywise error-rate to the nominal level. Here, we discuss an alternative method based on Monte Carlo simulation that allows the group size and stopping boundaries of a multi-arm multi-stage $t$-test to be optimised according to some nominated optimality criteria. We consider several examples, provide R code for general implementation, and show that our designs confer a familywise error-rate and power close to the desired level. Consequently, this methodology will provide utility in future multi-arm multi-stage trials.

**Keywords:** Familywise error-rate; Group sequential; Interim analyses; Multi-arm multi-stage; $t$-statistic.

## 1. Introduction

With the cost of drug development increasing, study designs that can enhance the efficiency of clinical research are of great interest. One such class of designs is the group sequential [1]. This approach exploits the fact that data are accumulated over time: incorporating interim analyses at which the study may be stopped early, reducing the required sample size.

Recently, this methodology was extended to allow multiple treatments to be compared to a shared control [2]. These multi-arm multi-stage (MAMS) designs can bring sizeable gains in efficiency over conducting a series of single-stage two-armed trials [3]. Unfortunately, a limitation of this methodology in the case of normally distributed outcome data is that designs are usually determined under the supposition of known patient variance in response. Typically, this will not be the case at the design stage. Then, utilising test statistics that assume known variance will result in operating characteristics that differ from their nominal level if the true variance is not equal to the specified value.

For two-armed group sequential trials, several authors have suggested methods to broach this problem. These include a recursive algorithm [4], and a quantile substitution procedure [1,5]. The latter approach was also explored for MAMS trials, and demonstrated to more accurately control the familywise error-rate (FWER) to the desired level, at a small cost to the trial's power [6].

A Monte Carlo based procedure was also proposed for two-armed group sequential trials [7]. In this paper, we extend it to MAMS trials. Explicitly, we describe how the stage-wise group size and stopping boundaries can be optimised. Finally, using the TAILoR trial [2] as a motivating example, we compare the performance of our method to several other approaches.

## 2. Methods

We consider a MAMS trial with $K+1$ arms, and a maximum of $J$ stages. Of the arms, $K$ (indexed $k = 1, \ldots, K$) are to be compared to a single control arm (indexed $k = 0$). We test the following hypotheses

$$H_0^{(k)} : \theta_k = \mu_k - \mu_0 \leq 0, \quad H_1^{(k)} : \theta_k = \mu_k - \mu_0 > 0, \quad k = 1, \ldots, K.$$

Here, $\mu_k$ is the mean response of patients allocated to arm $k = 0, \ldots, K$. We assume that in each stage, $n$ patients are allocated to each arm present in the trial. To allow for the early dropping of arms, we denote by $n_{kj}$ the actual number of patients allocated to arm $k = 0, \ldots, K$ in stage $j = 1, \ldots, J$. Thus, $n_{kj} \in \{0, n\}$. Designs with unequal allocation, or with two-sided null hypotheses could be treated similarly.

Denoting by $X_{kji}$ the response of the $i$th patient, in treatment arm $k$, in stage $j$, we assume that the $X_{kji}$ are independent and distributed as $X_{kji} \sim N(\mu_k, \sigma^2)$. Extending [7], set

$$N_{kj} = \sum_{l=1}^{j} n_{kl},$$

$$\bar{X}_{kj} = \frac{1}{N_{kj}} \sum_{l=1}^{j} \sum_{i=1}^{n} X_{kli},$$

$$\hat{\sigma}_j^2 = \frac{1}{\left(\sum_{k=0}^{K} N_{kj}\right) - (K+1)} \sum_{k=0}^{K} \sum_{l=1}^{j} \sum_{i=1}^{n} \left(X_{kli} - \bar{X}_{kj}\right)^2,$$

where $X_{kji} = 0 \, \forall i$ if $n_{kj} = 0$. At interim analysis $j$ the following test statistics are constructed

$$T_{kj}(\sigma) = \frac{\bar{X}_{kj} - \bar{X}_{0j}}{\sigma\sqrt{\frac{1}{N_{0j}} + \frac{1}{N_{kj}}}}, \quad k = 1, \ldots, K.$$

When $\sigma$ is assumed known, the $T_{kj}(\sigma)$ are together multivariate normal (henceforth, the $z$-test statistics). With $\sigma$ replaced by its estimate $\hat{\sigma}_j$, the joint distribution of the resulting $t$-test statistics, $T_{kj} = T_{kj}(\hat{\sigma}_j)$, does not have a simple form. It is this that makes the determination of stopping boundaries for use with $t$-test statistics difficult.

The parameters describing a MAMS design are then fully specified given efficacy and futility stopping boundaries $\boldsymbol{e} = (e_1, \ldots, e_J)^T \in \mathbb{R}^J$ and $\boldsymbol{f} = (f_1, \ldots, f_J)^T \in \mathbb{R}^J$, with $e_J = f_J$ to ensure the trial has at most $J$ stages.

We now consider two categories of MAMS design: one that terminates the entire trial as soon as any null hypothesis is rejected, and one that stops recruitment only to those arms for which the corresponding null hypothesis has been accepted or rejected. These two types of design have been referred to as including simultaneous and separate stopping respectively [8].

To describe our design, we introduce the vectors $\boldsymbol{\psi} = (\psi_1, \ldots, \psi_K)^T$ and $\boldsymbol{\omega} = (\omega_1, \ldots, \omega_K)^T$, where

- $\psi_k \in \{0,1\}$, with $\psi_k = 1$ if $H_0^{(k)}$ is rejected, and $\psi_k = 0$ otherwise;
- $\omega_k \in \{1, \ldots, J\}$, with $\omega_k = j$ if $j$ is the analysis at which $H_0^{(k)}$ is rejected, accepted, or the whole trial is stopped and no decision on $H_0^{(k)}$ is made.

Our MAMS $t$-test is then defined as follows

1. Set $\boldsymbol{\psi} = (\psi_1, \ldots, \psi_K)^T = \boldsymbol{\omega} = (\omega_1, \ldots, \omega_K)^T = (0, \ldots, 0)^T$ and $j = 1$.
2. Conduct stage $j$ of the trial, allocating $n$ patients to the control arm, and $n$ patients to each arm $k$ with $\omega_k = 0$.
3. Compute the $T_{kj}$.
4. For $k = 1, \ldots, K$
   4.1. If $T_{kl} \in (f_l, e_l]$ for $l = 0, \ldots, j-1$ (with the convention $T_{k0} \in (f_0, e_0]\ \forall k$)
      4.1.1. If $T_{kj} \geq e_j$ reject $H_0^{(k)}$ and set $\psi_k = 1, \omega_k = j$;
      4.1.2. If $T_{kj} < f_j$ accept $H_0^{(k)}$ and set $\omega_k = j$;
5. When using the simultaneous stopping rule, if $\sum_{k=1}^{K} \mathbb{I}\{\psi_k = 1\} = 0$ and $\sum_{k=1}^{K} \mathbb{I}\{\omega_k = 0\} > 0$, set $j = j+1$ and return to 2. Else stop the trial, and for each $k$ with $\omega_k = 0$, set $\omega_k = j$. When using the separate stopping rule, if $\sum_{k=1}^{K} \mathbb{I}\{\omega_k = 0\} > 0$, set $j = j+1$ and return to 2. Else stop the trial.

On trial completion, $\boldsymbol{\psi}$ and $\boldsymbol{\omega}$ then conform to their designations above.

We would like to ensure that the FWER, the probability of rejecting at least one true null hypothesis, is controlled to some level $\alpha$. There are several ways to define power in a multi-arm setting. Here, as in [2] we desire power of at least $1 - \beta$ to reject $H_0^{(1)}$ when $\theta_1 = \delta_1$ and $\theta_k = \delta_0$ for $k = 2, \ldots, K$. This is the so-called pairwise power for $H_0^{(1)}$ (see, e.g., [9]).

To this end, define

$$\Xi_{\text{sim}} = \left\{ (\boldsymbol{\omega}, \boldsymbol{\psi}) \in \{1, \ldots, J\}^K \times \{0,1\}^K : \text{if } \sum_{k=1}^{K} \mathbb{I}\{\omega_k \leq j\}\mathbb{I}\{\psi_k = 1\} \geq 1 \text{ then } \sum_{k=1}^{K} \mathbb{I}\{\omega_k > j\} = 0 \ \forall j \right\},$$

$$\Xi_{\text{sep}} = \{(\boldsymbol{\omega}, \boldsymbol{\psi}) \in \{1, \ldots, J\}^K \times \{0,1\}^K\}.$$

Here $\mathbb{I}\{x\}$ is the indicator function on event $x$. Furthermore, $\Xi_{\text{sim}}$ and $\Xi_{\text{sep}}$ represent the set of possible $\boldsymbol{\omega}, \boldsymbol{\psi}$ combinations when using the simultaneous and separate stopping rules respectively.

Then, for $\Xi \in \{\Xi_{\text{sim}}, \Xi_{\text{sep}}\}$ set according to the chosen stopping rule, take

$$\Xi_{\text{rej}} = \left\{ (\boldsymbol{\omega}, \boldsymbol{\psi}) \in \Xi : \sum_{k=1}^{K} \mathbb{I}\{\psi_k = 1\} > 0 \right\},$$

$$\Xi_1 = \{(\boldsymbol{\omega}, \boldsymbol{\psi}) \in \Xi : \mathbb{I}\{\psi_1 = 1\} = 1\}.$$

Here, $\Xi_{\text{rej}}$ and $\Xi_1$ are respectively the subsets of the $\Xi$ such that at least one null hypothesis, or $H_0^{(1)}$, is rejected.

Denoting the probability of a particular $(\boldsymbol{\omega}, \boldsymbol{\psi})$ combination on trial completion for a given vector of treatment effects $\boldsymbol{\theta} = (\theta_1, \ldots, \theta_K)^T$ by $\mathbb{P}(\boldsymbol{\omega}, \boldsymbol{\psi} \mid \boldsymbol{\theta})$, we specify our required operating characteristics as

$$\alpha_{\text{FWER}} = \sum_{(\boldsymbol{\omega}, \boldsymbol{\psi}) \in \Xi_{\text{rej}}} \mathbb{P}(\boldsymbol{\omega}, \boldsymbol{\psi} \mid \mathbf{0}) \leq \alpha, \tag{1}$$

$$1 - \beta_{\text{power}} = \sum_{(\boldsymbol{\omega}, \boldsymbol{\psi}) \in \Xi_1} \mathbb{P}(\boldsymbol{\omega}, \boldsymbol{\psi} \mid \boldsymbol{\delta}) \geq 1 - \beta,$$

where $\boldsymbol{\delta} = (\delta_1, \delta_0, \ldots, \delta_0)^T$.

Additionally, we optimise our choices of $\boldsymbol{n}$, $\boldsymbol{e}$, and $\boldsymbol{f}$. In theory, this could be achieved for almost any optimality criteria, with several sensible choices having been previously proposed (see, e.g., [10]). Here, we focus on minimising some weighted combination of the expected sample sizes (ESSs) when $\boldsymbol{\theta} = \mathbf{0}$ and $\boldsymbol{\theta} = \boldsymbol{\delta}$, and the maximal possible sample size; an approach that has in several trial design settings proved effective [5,11]. Note that

$$\text{ESS}(\boldsymbol{\theta}) = \sum_{(\boldsymbol{\omega},\boldsymbol{\psi})\in\Xi} n\left(\max_k \omega_k + \sum_{k=1}^{K} \omega_k\right) \mathbb{P}(\boldsymbol{\omega},\boldsymbol{\psi} \mid \boldsymbol{\theta}).$$

This could therefore, following [12], be achieved by identifying the $n$, $\boldsymbol{e}$, and $\boldsymbol{f}$ that minimise the following function

$$w_1 \text{ESS}(\mathbf{0}) + w_2 \text{ESS}(\boldsymbol{\delta}) + w_3 nJ(K+1)$$
$$+ P\left\{\mathbb{I}\{\alpha_{\text{FWER}} > \alpha\}\left(\frac{\alpha_{\text{FWER}} - \alpha}{\alpha}\right) + \mathbb{I}\{\beta_{\text{power}} > \beta\}\left(\frac{\beta_{\text{power}} - \beta}{\beta}\right)\right\}.$$

Here $P \in \mathbb{R}^+$ is a penalty for designs with undesirable operating characteristics, taken as the sample size required by a corresponding single-stage design. Moreover, the $w_i \in \mathbb{R} \cup \{0\}$, for $i = 1,2,3$, are weights given towards the desires to minimise the three included factors. Note that previous work suggests that designs that place all of their weight on one of the three factors (e.g., $w_1 = 1, w_2 = w_3 = 0$), will perform particularly badly for other choices of the weights [5,11]. It is therefore advisable to consider a range of options for the weights, and also to take $w_i \neq 0$, for $i = 1,2,3$.

Unfortunately, the complex joint distribution of the $T_{dl}$ prevents us from calculating the $\mathbb{P}(\boldsymbol{\omega},\boldsymbol{\psi} \mid \boldsymbol{\theta})$ required for this exactly. Instead, we use a Monte Carlo method. We offer first a more practical description of how this works, before providing a formal description below.

Suppose as an example that $J = K = 2$, $\mu_0 = \mu_1 = \mu_2 = 0$, $\sigma^2 = 1$, and that we will use the simultaneous stopping rule. For any choice of values for $n$, $\boldsymbol{e} = (e_1, e_2)^T$, and $\boldsymbol{f} = (f_1, f_2)^T$ (with $e_2 = f_2$), we can simulate a trials outcome by generating data from each treatment arm in stage one, using the fact that $X_{k1i} \sim N(0,1)$ for $k = 0,1,2$ and $i = 1, \dots, n$. With this data, the $T_{k1}$ for $k = 1, 2$ can be formed. If $T_{k1} \geq e_1$ for $k = 1$ or 2, the trial terminates, with a familywise error (FWE) having occurred. If $T_{k1} < f_1$ for $k = 1$ and 2, then the trial also terminates here, with no FWE having occurred. Otherwise the trial progresses to stage 2, with recruitment continued in arm 0 and the arms $k$ with $f_1 \leq T_{k1} < e_1$. We draw data for stage two in these arms again using the standard normal distribution, and then compute the $T_{k2}$ for those $k$ with $f_1 \leq T_{k1} < e_1$. The trial now terminates, either with a FWE having been committed if for at least one of these $k$, $T_{k2} \geq e_2$, or without a FWE having been committed otherwise. The FWER can then be estimated for this design by repeating the above process many times, and counting the proportion of instances in which a null hypothesis is rejected. Similarly, one can estimate the power under the LFC, or estimate ESSs. A global optimisation routine can then be used to search for the optimal values of $n$, $\boldsymbol{e}$, and $\boldsymbol{f}$.

Formally, we generate $R = 100{,}000$ independent sets of responses for each treatment arm under $\boldsymbol{\theta} = \boldsymbol{0}$ for some suitably large value of $n$. Subsets of these datasets are then used to form the responses for any smaller value of $n$. Next, for any $n$, $\boldsymbol{e}$, and $\boldsymbol{f}$, and chosen stopping rule, for the $r$th dataset, the trial is conducted as specified above. Importantly, the values of $\boldsymbol{\omega}$ and $\boldsymbol{\psi}$ on trial completion are determined, and denoted $\boldsymbol{\omega}_r$ and $\boldsymbol{\psi}_r$. An approximation to $\alpha_{\text{FWER}}$ for this design is then

$$\hat{\alpha}_{\text{FWER}} = \frac{1}{R} \sum_{r=1}^{R} \mathbb{I}\{(\boldsymbol{\omega}_r, \boldsymbol{\psi}_r) \in \Xi_{\text{rej}}\}.$$

We can similarly compute approximations $\hat{\beta}_{\text{power}}$, $\widehat{\text{ESS}}(\boldsymbol{0})$, and $\widehat{\text{ESS}}(\boldsymbol{\delta})$ to $\beta_{\text{power}}$ $\text{ESS}(\boldsymbol{0})$, and $\text{ESS}(\boldsymbol{\delta})$. Thus, to find the optimal design, we minimise the following function in $n$, $\boldsymbol{e}$ and $\boldsymbol{f}$

$$w_1 \widehat{\text{ESS}}(\boldsymbol{0}) + w_2 \widehat{\text{ESS}}(\boldsymbol{\delta}) + w_3 nJ(K+1)$$
$$+ P\left\{\mathbb{I}\{\hat{\alpha}_{\text{FWER}} > \alpha\}\left(\frac{\hat{\alpha}_{\text{FWER}} - \alpha}{\alpha}\right) + \mathbb{I}\{\hat{\beta}_{\text{power}} > \beta\}\left(\frac{\hat{\beta}_{\text{power}} - \beta}{\beta}\right)\right\}.$$

Note that the requirement to generate datasets necessitates $n$ to be treated as an integer. Thus, an algorithm that can simultaneously search over the discrete $n$, and the continuous $\boldsymbol{e}$ and $\boldsymbol{f}$ is required. We achieve this using CEoptim in R [13]. Code to implement our method is available from https://sites.google.com/site/jmswason/supplementary-material.

## 3. Results

We consider examples based on the TAILoR trial [2]. The trial tested three experimental treatments. We therefore take $K = 3$, and as an example set $J = 2$, $\sigma^2 = 1$, $\alpha = 0.05$, and $\beta = 0.1$. Conforming to our recommendations above, we additionally take $w_1 = w_2 = w_3 = 1/3$ (the 'balanced-optimal design'). As in previous work, we consider two scenarios [6]. For Scenario 1 we set $\delta_1 = 0.545$, $\delta_0 = 0.178$, and for Scenario 2, $\delta_1 = 1$, $\delta_0 = 0$.

For both scenarios, and both considered stopping rules, we determined the balanced-optimal design for $t$-test statistics using the Monte Carlo method described above (denoting the optimal values by $n_t$, $\boldsymbol{e}_t$, $\boldsymbol{f}_t$). For comparison, we use the triangular designs [14] for $z$-test statistics (denoting the values by $n_z$, $\boldsymbol{e}_z$, $\boldsymbol{f}_z$), which can be found using the MAMS package in R [2]. These designs are so-named for the shape of their stopping regions, can be found quickly, and have been shown to provide good performance in terms of their associated ESSs for MAMS trials [12]. The resultant designs are given in Table 1. Note that it is by construction of the triangular test that the boundaries are equal in each instance, subject to numerical error.

| Scenario | Stopping rule | Triangular Design | | | Balanced-Optimal Design | | |
|---|---|---|---|---|---|---|---|
| | | $n_z$ | $f_z$ | $e_z$ | $n_t$ | $f_t$ | $e_t$ |
| Scenario 1 | Simultaneous | 45 | $(0.777, 2.197)^T$ | $(2.330, 2.197)^T$ | 41 | $(0.606, 2.084)^T$ | $(2.742, 2.084)^T$ |
| Scenario 1 | Separate | 43 | $(0.777, 2.198)^T$ | $(2.330, 2.197)^T$ | 40 | $(0.721, 2.052)^T$ | $(2.925, 2.052)^T$ |
| Scenario 2 | Simultaneous | 13 | $(0.777, 2.197)^T$ | $(2.330, 2.197)^T$ | 12 | $(0.603, 2.010)^T$ | $(2.942, 2.010)^T$ |
| Scenario 2 | Separate | 13 | $(0.777, 2.197)^T$ | $(2.330, 2.197)^T$ | 12 | $(0.668, 2.086)^T$ | $(2.990, 2.086)^T$ |

**Table 1:** The triangular designs determined using the known variance test statistics, and the balanced-optimal designs determined using the unknown variance test statistics, are displayed for the two considered trial design scenarios, and the two considered stopping rules. All boundaries are given to three decimal places.

We then examined, using 100,000 trial simulations, the performance of the following approaches as a function of the true variance $\sigma_T^2$

A1. $n_z, e_z, f_z$ with $z$-test statistics and the presumed value of $\sigma^2$;
A2. $n_z, e_z, f_z$ with $t$-test statistics;
A3. $n_z, e_z, f_z$ with $t$-test statistics, and modification of the $e_z, f_z$ using quantile substitution. That is, at interim analysis $j$ we replace $e_{zj}$ and $f_{zj}$ by $e'_{zj} = T_{\sum_{k=0}^{K} N_{kj} - (K+1)} \left(1 - \Phi(e_{zj})\right)$ and $f'_{zj} = T_{\sum_{k=0}^{K} N_{kj} - (K+1)} \left(1 - \Phi(f_{zj})\right)$, where $T_\nu$ is the cumulative distribution function of Student's $t$-distibution with $\nu$ degrees of freedom;
A4. $n_t, e_t, f_t$ with the $t$-test statistics.

The results of these comparisons are given in Table 2. In both scenarios, using either stopping rule, assumption of known variance results in large inflation of the FWER when $\sigma_T^2 > \sigma^2$. In contrast, Approaches 3 and 4 far more accurately control the FWER in all cases, with Approach 4 controlling to the nominal level on slightly more occasions overall. Moreover, whilst $\widehat{\mathrm{ESS}}(\boldsymbol{\delta})$ is comparable for Approaches 3 and 4, Approach 4 always attains a lower value for $\widehat{\mathrm{ESS}}(\boldsymbol{0})$.

## 4. Discussion

In this article, we extended previous work for two-armed group sequential trials to allow the design parameters of a MAMS $t$-test to be optimised, when employing either a simultaneous or separate stopping rule. For the considered examples, the method was successful in providing operating characteristics close to their nominal level.

It is important to note that by Equation (1), the FWER is controlled under the global null hypothesis ($\boldsymbol{\theta} = \boldsymbol{0}$). This is known to provide strong control under the assumption of known variance with $z$-test statistics [2]. However, it is not known whether this is the case for the $t$-test statistics considered here. Therefore, whilst intuitively it seems logical that Equation (1) would provide strong control in this setting, a search over the vector $\boldsymbol{\theta}$ should be employed after initial design determination to verify this.

In conclusion, our method provides an alternative approach for dealing with unknown variance to the heuristic quantile substitution procedure. Precisely, quantile substitution offers a quick, often effective means of controlling the FWER relatively accurately. However, if it is vital to control the FWER, the proposed method should be preferable, and additionally allows the stopping boundaries to be optimised. In certain circumstances it can therefore be expected to allow the determination of more efficient designs.

| Factor | Approach | Scenario 1 $\sigma_T^2$ | | | | | Scenario 2 $\sigma_T^2$ | | | | |
|---|---|---|---|---|---|---|---|---|---|---|---|
| | | 0.25 | 0.5 | 1.0 | 2.0 | 4.0 | 0.25 | 0.5 | 1.0 | 2.0 | 4.0 |
| **Simultaneous stopping rule designs** | | | | | | | | | | | |
| $\hat{\alpha}_{\text{FWER}}$ | A1 | 0.0000 | 0.0035 | 0.0499 | 0.1816 | 0.3421 | 0.0000 | 0.0035 | 0.0495 | 0.1820 | 0.3450 |
| | A2 | 0.0508 | 0.0508 | 0.0518 | 0.0517 | 0.0514 | 0.0582 | 0.0561 | 0.0556 | 0.0570 | 0.0557 |
| | A3 | 0.0491 | 0.0492 | 0.0501 | 0.0497 | 0.0496 | 0.0519 | 0.0497 | 0.0500 | 0.0503 | 0.0495 |
| | A4 | 0.0493 | 0.0490 | 0.0504 | 0.0504 | 0.0487 | 0.0510 | 0.0487 | 0.0495 | 0.0494 | 0.0489 |
| $1-\hat{\beta}_{\text{power}}$ | A1 | 0.9981 | 0.9776 | 0.9078 | 0.7986 | 0.6949 | 0.9970 | 0.9740 | 0.9100 | 0.8120 | 0.7140 |
| | A2 | 1.0000 | 0.9952 | 0.9080 | 0.6314 | 0.3541 | 1.000 | 0.9960 | 0.9090 | 0.6330 | 0.3610 |
| | A3 | 0.9999 | 0.9951 | 0.9068 | 0.6276 | 0.3498 | 1.000 | 0.9960 | 0.9030 | 0.6180 | 0.3450 |
| | A4 | 0.9999 | 0.9939 | 0.9017 | 0.6258 | 0.3516 | 1.000 | 0.9960 | 0.9010 | 0.6210 | 0.3530 |
| $\widehat{\text{ESS}}(\mathbf{0})$ | A1 | 194.2 | 210.6 | 224.6 | 225.3 | 216.2 | 56.2 | 60.9 | 64.8 | 65.0 | 62.6 |
| | A2 | 223.8 | 224.0 | 224.4 | 224.3 | 224.0 | 64.7 | 64.7 | 64.7 | 64.6 | 64.8 |
| | A3 | 223.9 | 224.1 | 224.5 | 224.4 | 224.1 | 64.8 | 64.8 | 64.8 | 64.7 | 64.9 |
| | A4 | 216.0 | 216.1 | 216.7 | 216.3 | 216.2 | 63.6 | 63.5 | 63.5 | 63.5 | 63.6 |
| $\widehat{\text{ESS}}(\boldsymbol{\delta})$ | A1 | 216.4 | 222.1 | 222.6 | 217.6 | 208.8 | 60.6 | 62.0 | 62.6 | 61.9 | 60.2 |
| | A2 | 180.3 | 190.4 | 222.5 | 246.8 | 252.0 | 52.1 | 54.7 | 62.5 | 68.3 | 69.9 |
| | A3 | 180.3 | 190.9 | 223.6 | 247.8 | 252.8 | 52.1 | 55.2 | 63.4 | 69.2 | 70.5 |
| | A4 | 165.6 | 190.5 | 232.6 | 251.3 | 250.3 | 48.7 | 55.9 | 66.4 | 70.7 | 70.6 |
| **Separate stopping rule designs** | | | | | | | | | | | |
| $\hat{\alpha}_{\text{FWER}}$ | A1 | 0.0000 | 0.0035 | 0.0494 | 0.1820 | 0.3410 | 0.0000 | 0.0035 | 0.0507 | 0.1818 | 0.3461 |
| | A2 | 0.0509 | 0.0519 | 0.0519 | 0.0517 | 0.0522 | 0.0569 | 0.0561 | 0.0567 | 0.0575 | 0.0568 |
| | A3 | 0.0489 | 0.0500 | 0.0501 | 0.0499 | 0.0504 | 0.0501 | 0.0497 | 0.0504 | 0.0509 | 0.0499 |
| | A4 | 0.0494 | 0.0501 | 0.0497 | 0.0498 | 0.0508 | 0.0504 | 0.0498 | 0.0499 | 0.0506 | 0.0497 |
| $1-\hat{\beta}_{\text{power}}$ | A1 | 0.9970 | 0.9720 | 0.9060 | 0.8110 | 0.7260 | 0.9975 | 0.9747 | 0.9096 | 0.8168 | 0.7292 |
| | A2 | 1.0000 | 0.9960 | 0.9050 | 0.6220 | 0.3490 | 1.0000 | 0.9964 | 0.9080 | 0.6347 | 0.3625 |
| | A3 | 1.0000 | 0.9960 | 0.9040 | 0.6170 | 0.3440 | 1.0000 | 0.9960 | 0.9020 | 0.6183 | 0.3462 |
| | A4 | 1.0000 | 0.9950 | 0.9000 | 0.6220 | 0.3520 | 1.0000 | 0.9953 | 0.8992 | 0.6215 | 0.3536 |
| $\widehat{\text{ESS}}(\mathbf{0})$ | A1 | 185.6 | 201.2 | 217.0 | 224.1 | 222.5 | 56.1 | 60.9 | 65.5 | 67.8 | 67.3 |
| | A2 | 216.6 | 216.3 | 217.0 | 216.6 | 216.7 | 65.5 | 65.4 | 65.5 | 65.5 | 65.6 |
| | A3 | 216.5 | 216.3 | 217.0 | 216.6 | 216.7 | 65.5 | 65.4 | 65.5 | 65.5 | 65.6 |
| | A4 | 205.7 | 205.6 | 206.2 | 205.7 | 205.8 | 62.7 | 62.6 | 62.7 | 62.7 | 62.8 |
| $\widehat{\text{ESS}}(\boldsymbol{\delta})$ | A1 | 271.1 | 270.4 | 263.5 | 250.0 | 234.7 | 63.4 | 67.7 | 70.7 | 70.9 | 68.8 |
| | A2 | 253.5 | 255.8 | 263.3 | 263.3 | 255.3 | 61.8 | 64.2 | 70.7 | 73.9 | 73.2 |
| | A3 | 254.3 | 256.6 | 264.0 | 263.9 | 255.6 | 61.8 | 64.6 | 71.4 | 74.4 | 73.4 |
| | A4 | 254.1 | 257.9 | 263.9 | 257.9 | 245.9 | 59.4 | 65.6 | 72.1 | 73.0 | 71.0 |

**Table 2:** The estimated familywise error-rate ($\hat{\alpha}_{\text{FWER}}$), power ($1-\hat{\beta}_{\text{power}}$), and expected sample sizes (ESSs) when $\boldsymbol{\theta}=\mathbf{0}$ ($\widehat{\text{ESS}}(\mathbf{0})$) and $\boldsymbol{\theta}=\boldsymbol{\delta}$ ($\widehat{\text{ESS}}(\boldsymbol{\delta})$) of the four considered approaches (A1-A4) are shown as the true variance $\sigma_T^2$ varies, for the two considered trial design scenarios, and the two considered stopping rules. The rejection rate and ESS values are given to four and one decimal places respectively.

## Acknowledgements

This work was supported by the Medical Research Council [grant number MC_UP_1302/6 to M.J.G. and A.P.M.]; and the National Institute for Health Research Cambridge Biomedical Research Centre [grant number MC_UP_1302/4 to J.M.S.W.].